\documentclass[3p,times,twocolumn]{elsarticle}

\usepackage{ecrc}

\usepackage{amsmath}

\volume{00}

\firstpage{1}

\journalname{Nuclear Physics B Proceedings Supplement}

\runauth{J. Koponen et al.}

\jid{nuphbp}

\jnltitlelogo{Nuclear Physics B Proceedings Supplement}

\usepackage{amssymb}
\biboptions{comma, square}

\usepackage[figuresright]{rotating}

\begin{document}

\begin{frontmatter}



\dochead{}

\title{The strange and charm quark contributions to the anomalous magnetic moment of the muon from lattice QCD}


\author[label1]{Jonna Koponen\corref{cor1}}
\author[label1]{Bipasha Chakraborty}
\author[label1]{Christine T.~H. Davies}
\author[label2]{Gordon Donald}
\author[label3]{Rachel Dowdall}
\author[label1]{Pedro Gon\c calves de Oliveira}
\author[label4]{G. Peter Lepage}
\author[label5]{Thomas Teubner}

\cortext[cor1]{Speaker}

\address[label1]{SUPA, School of Physics and Astronomy, University of Glasgow, Glasgow, G12 8QQ, UK}
\address[label2]{Institut f\"ur Theoretische Physik, Universit\"at Regensburg, 93040 Regensburg, Germany}
\address[label3]{DAMTP, University of Cambridge, Wilberforce Road, Cambridge, CB3 0WA, UK}
\address[label4]{Laboratory for Elementary-Particle Physics, Cornell University, Ithaca, New York 14853, USA}
\address[label5]{Department of Mathematical Sciences, University of Liverpool, Liverpool, L69 3BX, UK}

\begin{abstract}

We describe a new technique (published in \cite{PhysRevD.89.114501}) to determine 
the contribution to the anomalous magnetic moment of the muon coming 
from the hadronic vacuum polarisation using lattice QCD. Our method uses 
Pad\'e approximants to reconstruct the Adler function from its derivatives 
at $q^2=0$. These are obtained simply and accurately from time-moments of 
the vector current-current correlator at zero spatial momentum. We test the 
method using strange quark correlators calculated on MILC Collaboration's 
$n_f = 2+1+1$ HISQ ensembles at multiple values of the lattice spacing, 
multiple volumes and multiple light sea quark masses (including physical 
pion mass configurations). We find the (connected) contribution to the 
anomalous moment from the strange quark vacuum polarisation to be 
$a^s_\mu=53.41(59)\times 10^{-10}$, and the contribution from charm quarks to be 
$a^c_\mu=14.42(39)\times 10^{-10}$ - 1\% accuracy is achieved for the strange 
quark contribution. The extension of our method to the light quark 
contribution and to that from the quark-line disconnected diagram is
straightforward.
\end{abstract}


\end{frontmatter}

% \linenumbers

\section{Motivation}

The magnetic moment of the muon can be determined extremely
accurately in experiment and the anomaly, $a_\mu = (g_\mu-2)/2$, is
known to 0.5 ppm (Brookhaven E821, \cite{PhysRevD.73.072003}).
Theoretical calculation of $a_\mu$ in the Standard Model shows a
discrepancy with the experimental result of about 
$25(9)\times 10^{-10}$, which could be an indication of new virtual
particles. Improvements of a factor of 4 in the experimental
uncertainty are expected in 2017-18 (Fermilab E989 experiment)
--- improvements in the theoretical determination would make the
discrepancy (if it remains) really compelling.

\begin{figure}
\centering
\includegraphics[width=0.24\textwidth]{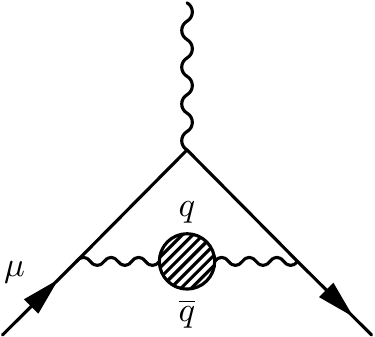}
\caption{Diagrammatic representation of the hadronic vacuum polarisation
contribution to the muon anomalous magnetic moment. The wavy lines are
photons and the shaded blob is the hadronic vacuum polarisation contribution
with all the quark and gluon interactions.}
\label{fig:hvpdiagram}
\end{figure}

The theoretical uncertainty is dominated by that from the hadronic vacuum
polarisation (HVP) contribution (Fig.~\ref{fig:hvpdiagram}). At the moment
the most accurate determination of HVP contribution comes from experiment:
dispersion relation + cross section for $e^+e^-$ (and $\tau$) $\to$ hadrons
gives $\sim 700\times 10^{-10}$  with a 1\% error 
\cite{0954-3899-38-8-085003, EPJC71_1515}. 
Higher order contributions from QCD processes have
larger percentage uncertainty but make an order of magnitude smaller
contribution, so we set out to calculate the lowest order HVP contribution
and aim at $1-2$\% accuracy.

\section{Method}

We express the $a_\mu$ HVP contribution in terms of a small number of derivatives
of the hadronic vacuum polarisation function $\Pi$ evaluated at zero momentum and
use lattice current-current correlators to calculate the derivatives
\cite{PhysRevD.89.114501}.

The HVP contribution to the anomalous magnetic moment for quark flavour $\mathrm{f}$ is
\begin{equation}
a_{\mu, \mathrm{HVP}}^{(\mathrm{f})}=\frac{\alpha}{\pi}\int_0^\infty
\!\!\!\mathrm{d}q^2 f(q^2)\big(4\pi\alpha Q^2_{\mathrm{f}}\big)\hat{\Pi}_{\mathrm{f}}(q^2),
\end{equation}
where
\begin{equation}
\begin{split}
f(q^2)&\equiv \frac{m^2_\mu q^2A^3(1-q^2A)}{1+m^2_\mu q^2A^2},\\
A&\equiv \frac{\sqrt{q^4+4m^2_\mu q^2}-q^2}{2m^2_\mu q^2}
\end{split}
\end{equation}
and $\alpha=\alpha_{\mathrm{QED}}$. The integrand peaks at 
$q^2\sim \mathcal{O}(m^2_\mu)$. Previous methods calculated 
$\hat{\Pi}(q^2)$ at larger $q^2$ and extrapolated to zero, which
leads to model uncertainties, and calculating directly at small $q^2$
using ``twisted boundary conditions'' produces noisy results. We avoid
this by working, in effect, from $q^2=0$ upwards.

The vacuum polarisation function is a Fourier transform of the lattice
current-current correlator:
\begin{equation}
\Pi^{ii}=q^2\Pi(q^2)=a^4\sum_t \mathrm{e}^{iqt}\sum_{\vec{x}}
\langle j^i(\vec{x},t)j^i(0)\rangle.
\end{equation}
We calculate the derivatives of the renormalised vacuum polarisation function
$\hat{\Pi}(q^2)\equiv \Pi(q^2)-\Pi(0)$ from the time moments of the correlator:
\begin{equation}
\begin{split}
G_{2n}&\equiv a^4\sum_t\sum_{\vec{x}}t^{2n}Z^2_V\langle j^i(\vec{x},t)j^i(0)\rangle\\
&=(-1)^n\frac{\partial^{2n}}{\partial q^{2n}}q^2\hat{\Pi}(q^2)\Big|_{q^2=0}.
\end{split}
\end{equation}
Here the correlator $\langle j^i(\vec{x},t)j^i(0)\rangle$ is a local
spatial vector current at zero spatial momentum.

We define $\hat{\Pi}(q^2)$ through the series expansion
\begin{equation}
\hat{\Pi}(q^2)=\sum_{J=1}^{\infty}q^2\Pi_j,\; \Pi_j=(-1)^{j+1}\frac{G_{2j+2}}{(2j+2)!}
\end{equation}
and use Pad\'e approximants for $\hat{\Pi}(q^2)$ to control the high-$q^2$ region.
High-order Pad\'e approximants converge to the exact result~\cite{PhysRevD.89.114501}.
We use 4th, 6th, 8th and 10th time moments (i.e. $j=1$, 2, 3 and 4). Only
quark-line connected contributions to the lowest order HPV are considered here
--- disconnected contributions will need to be addressed separately.

\section{Lattice configurations}

We use lattice ensembles made by MILC collaboration 
\cite{PhysRevD.82.074501, PhysRevD.87.054505}
that have $u/d$, $s$ and $c$ quarks in the sea (i.e.
$n_f = 2+1+1$). The lattice spacings are $a\approx 0.15$~fm (very coarse),
$0.12$~fm (coarse) and $0.09$~fm (fine), determined using the Wilson flow parameter $w_0$
\cite{PhysRevD.88.074504}. We use Highly Improved Staggered  Quark (HISQ)
action, which is known for very small discretisation errors. We use ensembles
that have different light sea quark masses, including ensembles with physical
light quark mass. The strange valence quark is tuned using the $\eta_s$ meson
mass $m_{\eta_s}=688.5$~MeV \cite{PhysRevD.88.074504}. We also test tuning effects
by deliberately mistuning the strange quark by 5\% (set 6). We use large volumes,
$(5.6 \textrm{ fm})^3$ on the finest lattices, and have ensembles with different
volumes for testing finite volume effects. Details of the lattice ensembles are
listed in Table~\ref{tbl:ensembles}.

The local current used here is not the conserved vector current for this quark
action and must be normalised. Renormalisation constant $Z_{V,\bar{s}s}$ is calculated
completely nonperturbatively by demanding that the vector form factor for this
current be 1 between two equal mass mesons at rest ($q^2 = 0$)
\cite{Chakraborty:2014zma}. Accurate normalisation is crucial if the target is
total uncertainty at the $1$\% level.

\section{Meson correlators}

\begin{figure}
\centering
\includegraphics[angle=-90,width=0.42\textwidth]{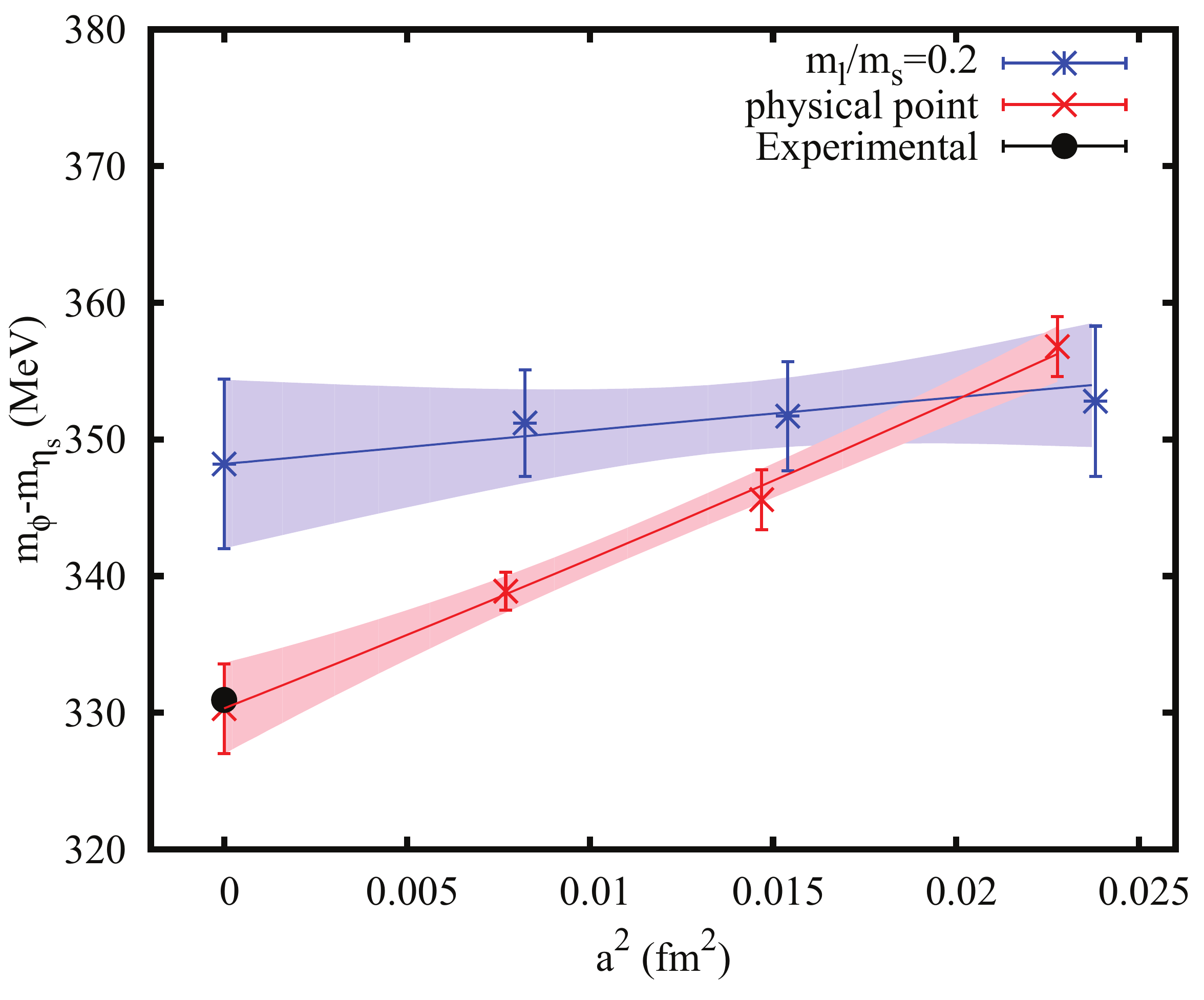}
\caption{Mass difference $m_\phi-m_{\eta_s}$ as a function of $a^2$.
Comparing results on ensembles that have different light sea quark masses
shows the advantage of working at the physical point.}
\label{fig:mphi}
\end{figure}

\begin{figure}
\centering
\includegraphics[angle=-90,width=0.435\textwidth]{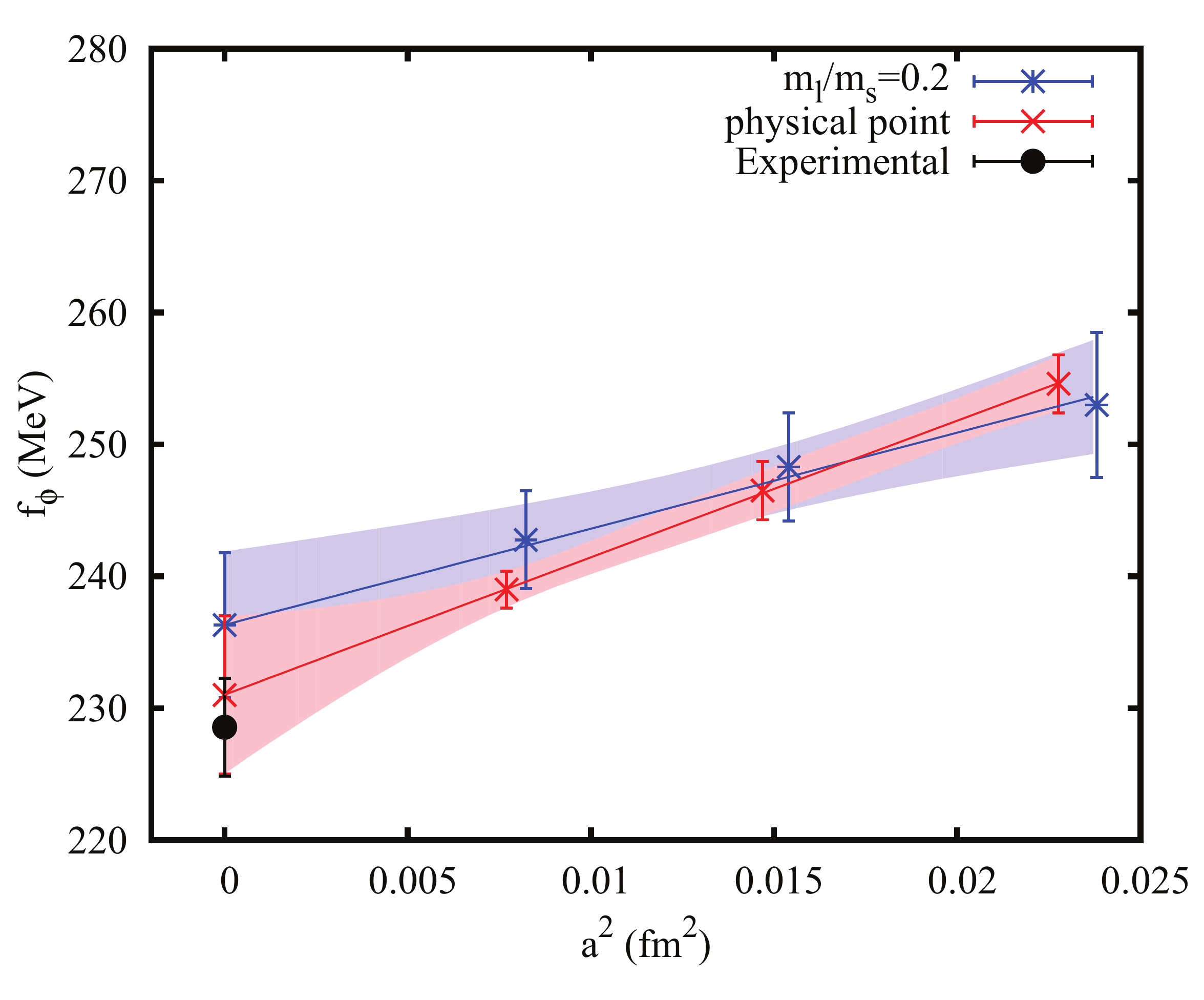}
\caption{$\phi$ meson decay constant as a function of $a^2$,
comparing results on ensembles that have different light sea quark masses
($m_l^{\mathrm{lat}}=m_s/5$ and $m_l^{\mathrm{lat}}=m_l^{\mathrm{phys}}$).}
\label{fig:fphi}
\end{figure}

The 2-point correlators used in this study are the $\phi$ meson correlators
made using a local spatial vector operator. As a cross-check and illustration
of the accuracy of the correlators we plot the mass difference of the $s\bar{s}$
vector and pseudoscalar mesons, $m_\phi-m_{\eta_s}$, in Fig.~\ref{fig:mphi}, and
also plot the $\phi$ meson decay constant in Fig.~\ref{fig:fphi}. Both figures
show that the discretisation errors are indeed small and also emphasize the
importance of working at the physical light quark masses (here sea quarks,
but even more important in the case of light valence quarks).

\begin{figure}
\centering
\includegraphics[width=0.435\textwidth]{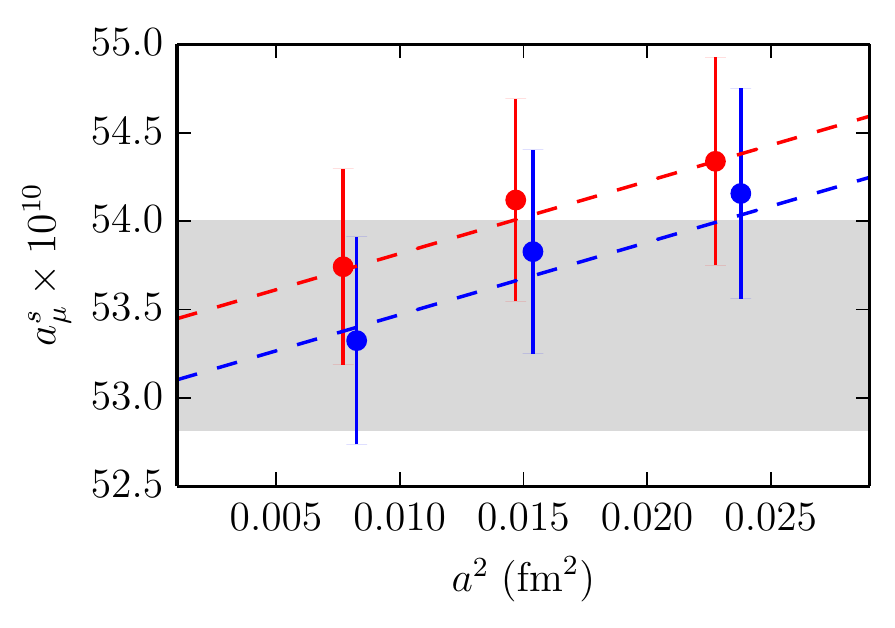}
\caption{Continuum extrapolation fit. Here we compare results from
ensembles with $m_l^{\mathrm{lat}}=m_s/5$ (blue points) to results from
ensembles with physical light quark mass (red points). The shaded error
band is our final result.}
\label{fig:amufit}
\end{figure}

\section{Fitting the data}

We use the [2,2] Pad\'e approximant for each configuration set and then fit
these results to a function of the form
\begin{equation}
a^s_{\mu, \mathrm{lat}}=a^s_\mu \Bigg(1+c_{a^2}\bigg(\frac{a\Lambda_{\mathrm{QCD}}}{\pi}\bigg)^2
+c_{\mathrm{sea}}\delta_{\mathrm{sea}}+c_{\mathrm{val}}\delta_{\mathrm{val}}\Bigg),
\end{equation}
where $\Lambda_{\mathrm{QCD}}=0.5$~GeV and
\begin{equation}
\delta_{\mathrm{sea}}\equiv \sum_{q=u,d,s}\frac{m^{\mathrm{sea}}_q
-m^{\mathrm{phys}}_q}{m^{\mathrm{phys}}_s},\;
\delta_{\mathrm{val}}\equiv \frac{m^{\mathrm{val}}_s
-m^{\mathrm{phys}}_s}{m^{\mathrm{phys}}_s}.
\end{equation}
Here the fit parameter $c_{a^2}$ takes care of the (small) discretisation effects
and $c_{\mathrm{sea}}$ and $c_{\mathrm{val}}$ take care of the dependence on the sea
and valence quark masses. The fit results along with lattice data are shown in
Fig.~\ref{fig:amufit}. More details of the fit can be found in \cite{PhysRevD.89.114501}.

\section{Results}

Our result for the leading order HVP contribution to the muon anomalous
magnetic moment from the strange quarks is
$a_\mu^s = 53.41(59)\times 10^{-10}$ (connected pieces only). We also used
time moments we calculated in \cite{Davies:2013ju, PhysRevD.86.094501} to get
the charm quark contribution to $a_\mu$: $a_\mu^c=14.42(39)\times 10^{-10}$
(again, connected pieces only). A preliminary estimate of total
light, strange and charm quark connected piece contributions
(averaging $a_\mu^{\mathrm{light}}$ on very coarse and coarse physical $m_l$
ensembles) is
%\begin{equation}
$a_\mu^{\mathrm{HVP},\mathrm{LO}}=a_\mu^{\mathrm{light}}+a_\mu^s+a_\mu^c \sim 662(35)\times 10^{-10}$.
%\end{equation}
This is to be compared with the dispersion relation + $e^+e^- \to$ hadrons
cross section result of $\sim 700\times 10^{-10}$
\cite{0954-3899-38-8-085003, EPJC71_1515} mentioned earlier. Note that our
result does not include the disconnected diagrams.

The error budget for both strange and charm quark connected contributions is
given in Table~\ref{tab:errbudget}. The dominant error in $a_\mu^s$ is, by far,
that coming from the uncertainty in the physical value of the Wilson flow
parameter $w_0$, which we use to set the lattice spacings. The next largest
contribution comes from the uncertainty in the renormalisation factor $Z_V$.
For the charm quark contribution this is the dominant error, because a
different method for calculating $Z_V$ was used in that calculation. This
could be improved by using the same method that was used here for the strange
quark contribution. More details about the error budget are
in~\cite{PhysRevD.89.114501}.

\begin{figure}
\centering
\includegraphics[width=0.45\textwidth]{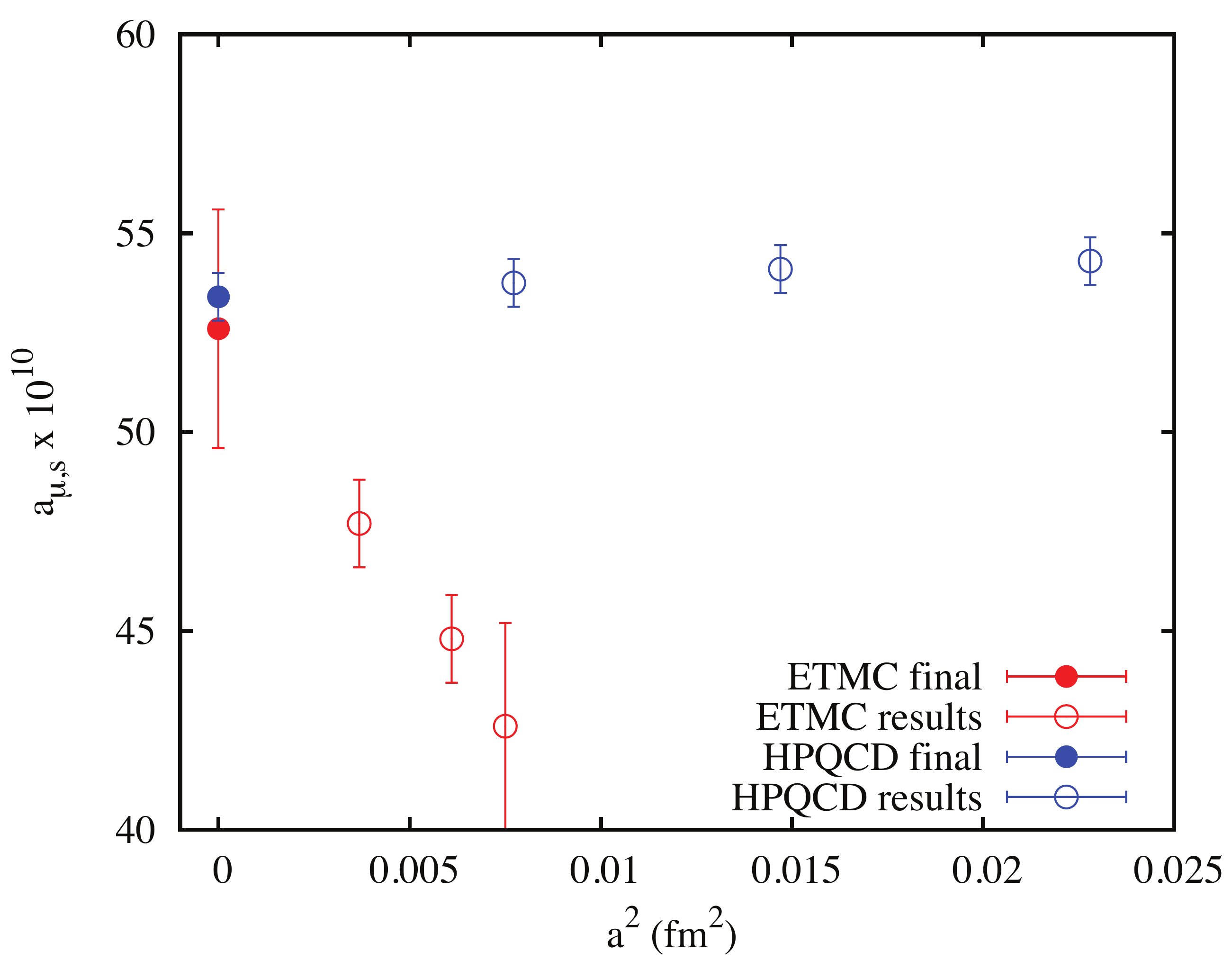}
\caption{Comparison with other lattice results: $a_\mu^s$ as a function of $a^2$.
ETMC results are from~\cite{Burger:2013jya}.}
\label{fig:amu_HPQCD_ETMC}
\end{figure}

Comparing our results with other lattice QCD results shows good agreement:
Fig.~\ref{fig:amu_HPQCD_ETMC} shows our results (HPQCD) and European Twisted
Mass Collaboration's results plotted against $a^2$. This again
highlights the fact that HISQ action has very small discretisation errors
compared to other lattice actions. RBC/UKQCD have also calculated $a_\mu^s$ and agree
well with other determinations --- see Table~\ref{tbl:comp} for a list of
the results.

\section{Connected contribution to $a_\mu^{\mathrm{light}}$}

%\begin{figure}
%\centering
%\includegraphics[width=0.475\textwidth]{amu.pdf}
%\caption{PRELIMINARY results: connected contributions to $a_\mu^{\mathrm{HVP}}$
%from charm, strange and light quarks plotted against the pseudoscalar
%meson mass (i.e. from left to right: light, strange and charm).}
%\label{fig:amulsc}
%\end{figure}

The signal-to-noise ratio at large $t$ is much worse for light quarks than for
strange or charm quarks, which means we will need better statistics and
improvements in the fitting method to get the error down to the 1\% level. We can
get better accuracy by calculating moments from best fit parameters instead
of using raw lattice data. 5-6\% precision is already achieved using 1000
configurations $\times$ 12 time sources (very coarse ensemble) and 400
configurations $\times$ 4 time sources (coarse ensemble) on physical
light quark mass ensembles. 
1\% precision can be achieved by adding more time sources
(need 4 times $n_{\mathrm{src}}$) and up to 10 $\times$ configurations.

\section{Conclusions}

We have demonstrated in~\cite{PhysRevD.89.114501} that 1\% precision can
be achieved for the leading order HVP contribution to $a_\mu^s$ from the
connected pieces. The error on $a_\mu^c$ could be pushed down to 1\% by
using the same method to calculate the renormalisation factor $Z_V$ that
was used here for the strange quark. However, the charm quark contribution
to the total leading order HVP contribution $a_\mu^{\mathrm{HVP},\mathrm{LO}}$
is small compared to contributions from strange and light quarks so this
is not a top priority. The main task now is to push down the error coming from
the light quark contribution $a_\mu^l$. We can get good enough statistics to
achieve this:
we can use more time sources, plus more very coarse and
coarse configurations can be made relatively cheaply. Disconnected
contributions need to be included in the future.

\begin{table*}
\centering
\begin{tabular}{rcclccr}
\hline\hline
Set & $am^{\mathrm{sea}}_l$ & $am^{\mathrm{sea}}_s$ & \;\;\;\;$am_{\eta_s}$
& $Z_{V,\bar{s}s}$ & $L/a\times T/a$ & $n_{\mathrm{cfg}}\times n_{\mathrm{src}}$\\
\hline
1 & 0.01300 & 0.0650 & 0.54024(15)& 0.9887(20) & $16\times 48$ & $1020\times 12$ \\
2 & 0.00235 & 0.0647 & 0.52680(8) & 0.9887(20) & $32\times 48$ & $1000\times 12$ \\
\hline
3 & 0.01020 & 0.0509 & 0.43138(12)& 0.9938(17) & $24\times 64$ & $526 \times 16$ \\
4 & 0.00507 & 0.0507 & 0.42664(9) & 0.9938(17) & $24\times 64$ & $1019\times 16$ \\
5 & 0.00507 & 0.0507 & 0.42637(6) & 0.9938(17) & $32\times 64$ & $988 \times 16$ \\
6 & 0.00507 & 0.0507 & 0.41572(14)& 0.9938(17) & $32\times 64$ & $300 \times 16$ \\
7 & 0.00507 & 0.0507 & 0.42617(9) & 0.9938(17) & $40\times 64$ & $313 \times 16$ \\
8 & 0.00184 & 0.0507 & 0.42310(3) & 0.9938(17) & $48\times 64$ & $1000\times 16$ \\
\hline
9 & 0.00740 & 0.0370 & 0.31384(9) & 0.9944(10) & $32\times 48$ & $504 \times 16$ \\
10& 0.00120 & 0.0363 & 0.30480(4) & 0.9944(10) & $64\times 96$ & $621 \times 16$ \\
\hline\hline
\end{tabular}
\caption{Lattice ensembles used in this study, made by MILC collaboration
\cite{PhysRevD.82.074501, PhysRevD.87.054505}. The first two sets are ``very coarse''
(lattice spacing $a\sim 0.15$~fm), sets $3-8$ are ``coarse'' ($a\sim 0.12$~fm) 
and sets $9-10$ are ``fine'' ($a\sim 0.09$~fm) ensembles.  $am^{\mathrm{sea}}_l$ 
and $am^{\mathrm{sea}}_s$ are the sea light and strange quark masses in lattice 
units and $am_{\eta_s}$ is the $\eta_s$ meson mass.  $Z_{V,\bar{s}s}$ is the vector 
current renormalisation constant. $L$ and $T$ are the spatial and temporal 
extents of the lattice. $n_{\mathrm{cfg}}$ is the number of configurations and 
$n_{\mathrm{src}}$ is the number of time sources used in this study.}
\label{tbl:ensembles}
\end{table*}

\begin{table*}
\centering
\begin{tabular}{rcc}
& $a_\mu^s$ & $a_\mu^c$ \\
\hline
Uncertainty in lattice spacing ($w_0$, $r_1$): & 1.0\% & 0.6\% \\
Uncertainty in $Z_V$: & 0.4\% & 2.5\% \\
Monte Carlo statistics: & 0.1\% & 0.1\% \\
$a^2\to0$ extrapolation: & 0.1\% & 0.4\% \\
QED corrections: & 0.1\% & 0.3\% \\
Quark mass tuning: & 0.0\% & 0.4\% \\
Finite lattice volume: & $<0.1\%$ & 0.0\% \\
Pad\'e approximants: & $<0.1\%$ & 0.0\% \\
\hline
Total: & 1.1\% & 2.7\% 
\end{tabular}   
\caption{Error budgets for connected contributions to 
the muon anomaly~$a_\mu$ from vacuum polarization of $s$ and $c$ quarks.
See~\cite{PhysRevD.89.114501} for more detailed discussion on the estimation
of the errors.}
\label{tab:errbudget}
\end{table*}

\begin{table*}
\centering
\begin{tabular}{ccccc}
\hline\hline
$a_\mu^{s/c}$ & dispersion & HPQCD & ETMC & RBC/UKQCD \\
 & + expt & & (prelim.) & (prelim.) \\
\hline
$a_\mu^s \times 10^{10}$ & 55.3(8) & 53.4(6) & 53(3) & 52.4(2.1)\\
\hline
$a_\mu^c \times 10^{10}$ & 14.4(1) & 14.4(4) & 14.1(6) & -- \\
\hline\hline
\end{tabular}
\caption{Comparison with other results. The dispersion relation + experiment
results are from \cite{0954-3899-38-8-085003} and \cite{PhysRevD.85.014029};
HPQCD results are from \cite{PhysRevD.89.114501} (moments used for $a_\mu^c$
were calculated in \cite{Davies:2013ju, PhysRevD.86.094501}); ETMC results
are from \cite{Burger:2013jya};
RBC/UKQCD results are from \cite{MaltmanLat2014}.}
\label{tbl:comp}
\end{table*}

\section*{Acknowledgements}

We are grateful to the MILC collaboration for the use of their gauge configurations. 
Our calculations were done on the Darwin Supercomputer as part of STFC’s DiRAC
facility jointly funded by STFC, BIS and the Universities of Cambridge and Glasgow.



%% References

\nocite{*}
\bibliographystyle{elsarticle-num}
\bibliography{gmurefs}

\begin{thebibliography}{10}
\expandafter\ifx\csname url\endcsname\relax
  \def\url#1{\texttt{#1}}\fi
\expandafter\ifx\csname urlprefix\endcsname\relax\def\urlprefix{URL }\fi
\expandafter\ifx\csname href\endcsname\relax
  \def\href#1#2{#2} \def\path#1{#1}\fi

\bibitem{PhysRevD.89.114501}
B.~Chakraborty, C.~T.~H. Davies, G.~C. Donald, R.~J. Dowdall, J.~Koponen, G.~P.
  Lepage, T.~Teubner, Strange and charm quark contributions to the anomalous
  magnetic moment of the muon, Phys. Rev. D 89 (2014) 114501.

\bibitem{PhysRevD.73.072003}
G.~W. Bennett, et~al., Final report of the {E821} muon anomalous magnetic
  moment measurement at {BNL}, Phys. Rev. D 73 (2006) 072003.

\bibitem{0954-3899-38-8-085003}
K.~Hagiwara, R.~Liao, A.~D. Martin, D.~Nomura, T.~Teubner, {$(g − 2)_\mu$}
  and {$\alpha (M_Z^2)$} re-evaluated using new precise data, Journal of
  Physics G: Nuclear and Particle Physics 38~(8) (2011) 085003.

\bibitem{EPJC71_1515}
M.~Davier, A.~Hoecker, B.~Malaescu, Z.~Zhang, Reevaluation of the hadronic
  contributions to the muon {$g−2$} and to {$\alpha (M^{2}_{Z})$}, The
  European Physical Journal C 71~(1).

\bibitem{PhysRevD.82.074501}
A.~Bazavov, et~al., Scaling studies of {QCD} with the dynamical highly improved
  staggered quark action, Phys. Rev. D 82 (2010) 074501.

\bibitem{PhysRevD.87.054505}
A.~Bazavov, et~al., Lattice {QCD} ensembles with four flavors of highly
  improved staggered quarks, Phys. Rev. D 87 (2013) 054505.

\bibitem{PhysRevD.88.074504}
R.~J. Dowdall, C.~T.~H. Davies, G.~P. Lepage, C.~McNeile, ${V}_{us}$ from $\pi$
  and {$K$} decay constants in full lattice {QCD} with physical $u$, $d$, $s$,
  and $c$ quarks, Phys. Rev. D 88 (2013) 074504.

\bibitem{Chakraborty:2014zma}
B.~Chakraborty, C.~T.~H. Davies, G.~Donald, R.~Dowdall, J.~Koponen, G.~P.
  Lepage, {Nonperturbative tests of the renormalization of mixed
  clover-staggered currents in lattice {QCD}}, PoS LATTICE2013 (2013) 309.
\newblock \href {http://arxiv.org/abs/1401.0669} {\path{arXiv:1401.0669}}.

\bibitem{Davies:2013ju}
C.~Davies, G.~Donald, R.~Dowdall, J.~Koponen, E.~Follana, et~al., {Precision
  tests of the {$J/\psi$} from full lattice QCD: mass, leptonic width and
  radiative decay rate to {$\eta_c$}}, PoS ConfinementX (2012) 288.
\newblock \href {http://arxiv.org/abs/1301.7203} {\path{arXiv:1301.7203}}.

\bibitem{PhysRevD.86.094501}
G.~C. Donald, C.~T.~H. Davies, R.~J. Dowdall, E.~Follana, K.~Hornbostel,
  J.~Koponen, G.~P. Lepage, C.~McNeile, Precision tests of the {$J/\psi$} from
  full lattice {QCD}: Mass, leptonic width, and radiative decay rate to
  {$\eta_c$}, Phys. Rev. D 86 (2012) 094501.

\bibitem{Burger:2013jya}
F.~Burger, X.~Feng, G.~Hotzel, K.~Jansen, M.~Petschlies, D.~B. Renner,
  Four-flavour leading-order hadronic contribution to the muon anomalous
  magnetic moment, JHEP 02 (2014) 099, (see also note in Phys. Rev. D89 (2014)
  114501: private communication for the separate contributions {$a_\mu^s$} and
  {$a_\mu^c$}).

\bibitem{PhysRevD.85.014029}
S.~Bodenstein, C.~A. Dominguez, K.~Schilcher, Hadronic contribution to the muon
  {$g-2$} factor: A theoretical determination, Phys. Rev. D 85 (2012) 014029.

\bibitem{MaltmanLat2014}
K.~Maltman, A new strategy for evaluating the {LO} {HVP} contribution to
  {$(g-2)_\mu$} on the lattice, talk at the 32nd International Symposium on
  Lattice Field Theory {(Lattice 2014)} (June 2014).

\end{thebibliography}






% \end{thebibliography}

\end{document}